# Controlling frustrated liquids and solids with an applied field in a kagome Heisenberg antiferromagnet

Satoshi Nishimoto[1], Naokazu Shibata[2] & Chisa Hotta[3]*

[1] Institute for Theoretical Solid State Physics, IFW Dresden, 01171 Dresden, Germany
[2] Department of Physics, Tohoku University, Sendai 980-8578, Japan
[3] Department of Physics, Faculty of Science, Kyoto Sangyo University, Kyoto 603-8555, Japan.

Quantum spin-1/2 kagome Heisenberg antiferromagnet is the representative frustrated system possibly hosting a spin liquid. Clarifying the nature of this elusive topological phase is a key challenge in condensed matter, however, even identifying it still remains unsettled. Here, we apply a magnetic field and discover a series of spin gapped phases appearing at five different fractions of magnetization by means of grand canonical density matrix renormalization group, an unbiased state-of-art numerical technique. The magnetic field dopes magnons and first gives rise to a possible $Z_3$ spin liquid plateau at 1/9-magnetization. Higher field induces a self-organized super-lattice-unit, a six-membered ring of quantum spins, resembling an atomic orbital structure. Putting magnons into this unit one by one yields three quantum solid plateaus. We thus find that the magnetic field could control the transition between various emergent phases by continuously releasing the frustration.

It is widely accepted that condensed matter orders at low temperatures by spontaneously breaking some sort of symmetry – translational symmetry in crystalline solids, time reversal and rotational symmetries in magnets, gauge symmetry in superconductors, and so on. Whether they can escape from ordering and instead bear emergent states is a question that recurred over decades. A possible strategy to remove trivial orders is to design a model with fine balance of microscopic interactions, where the low energy states are "frustrated" and a macroscopic number of quasi-degenerate states compete with each other. When quantum fluctuations between these states prevent a selection of particular order, one ends up with quantum disordered spin liquids. Modern theories have brought us new insight by identifying spin liquids as topological phases of matter[1,2]. Even if such topological phases are not formed, the resultant phase would also be nontrivial; its smallest disentangled unit could consist of several degrees of freedom --- pairs of spins in spin ladders or valence bond solids, etc. Thus, characterizing the regime of such liquids and relevant nontrivial phases is now becoming an important challenge.

In reality, spin liquids are quite elusive[3]. There are experimental studies on the several materials suggesting the existence of a spin liquid, $\kappa$-(BEDT-TTF)$_2$Cu$_2$(CN)$_3$[4], BaCu$_3$V$_2$O$_8$(OH)$_2$[5] and ZnCu$_3$(OD)$_6$Cl$_2$[6]. In theories, a few candidates include a triangular lattice Mott insulator[7] and spin-1/2 kagome antiferromagnet[8-11], represented by the Hubbard and the Heisenberg models, respectively. However, a theoretical information directly compared to the experimental data is still severely lacking.

Therefore, to find realistic theoretical models that could realize several nontrivial phases, and could provide information relevant to experiments is essential. The phase transitions should better be controlled by the experimentally tunable parameter.

Here, we show that quantum spin-1/2 kagome Heisenberg antiferromagnet in an applied field could be an ideal playground providing numbers of exotic quantum liquid and solid phases under field-control. The model itself is already known as a representative frustrated system that embodies fine balance of interaction by use of the lattice geometry, and more importantly, it is experimentally relevant. It is defined by the following Hamiltonian:

$$\mathcal{H} = \sum_{\langle i,j \rangle} J S_i \cdot S_j - H \sum_{i=1}^{N} S_i^z$$

where $S_i$ is a spin-1/2 operator at the $i$-th lattice site ($S_i^z$ is the z-component), and $J$ and $H$ are the interaction coupling constant and the external magnetic field, respectively. The first summation denoted by $\langle i,j \rangle$ runs over nearest neighbour pairs of sites of the kagome lattice of size $N$. The ground state at $H=0$ is possibly a spin liquid, but its detailed identification still suffers numerical difficulty.

A key numerical problem in quantum many body models on two-dimensional frustrated lattices is the lack of method that affords sizable results. The quantum Monte Carlo method suffers from the sign problem, and the exact diagonalization requires unbiased size scaling, which is unavailable at present. Regarding the kagome antiferromagnet, The multiscale entanglement renormalization supported a valence bond crystal ground state with a hexagonal unit cell of 36 spins[11], which was taken over by a $Z_2$ gapped spin liquid lower in energy in the latest two-dimensional density matrix renormalization group (DMRG) studies on a long cylinder[8,9]. In addition, there is a recent study predicting a gapless $U(1)$ liquid[7] and the issue remains unsettled.

Clarifying the nature of the model in an applied field also demands a severe numerical challenge; the magnetization process appears not as a curve but as a staircase of height $\sim 1/N$ due to finite size effect. Thus, what is known so far is the highly possible presence of a spin gapped phase called plateau at a 1/3-magnetization[12-14] and a jump of the magnetization from 7/9 to 1 at the saturation field[15,16]. While, even a 1/3-plateau is really a plateau[12,13] or something else[14] was not really concluded.

In the present article, we determine the bulk magnetization process of the kagome Heisenberg antiferromagnet by means of a grand canonical analysis[17,18]. The system turns out to have five different plateau phases, and the magnetic field controls the successive phase transitions between these plateau phases and the gapless liquids in a strikingly analogous manner to the Hall conductivity of the quantum Hall effect[19-21]. Two of the plateaus at lower fields are the possible spin liquids, which are characterized by the finite topological dimensions. The latter three at higher fields form a long range order in a particular unit, a hexagonal plaquette consisting of six-membered ring of spins. This unit resembles a quantum mechanical atomic orbital that accommodates several magnons in its discrete energy levels. The magnetic field thus transforms the system from the highly frustrated liquid phases to the moderately frustrated solid ones.

## Results

**Magnetization curve.** To overcome the numerical finite size effect, we apply the grand canonical DMRG, which was developed very recently[17], and was successfully applied to two dimension[18]. This method gives us a numerically *exact and unbiased* magnetization *curve* in the thermodynamic limit (See Methods); For example, if we choose the cylinder of a finite circumference and of length $L$ which is larger than ~10 lattice spacings, and perform a grand canonical analysis, we obtain a magnetization curve of an infinitely long cylinder of that fixed circumference(See supplementary Fig.S1). Since we need a result of a bulk system, equivalently spanned along three different directions of a kagome lattice, we choose a hexagonal cluster for the present calculation, rather than a long cylinder.

Figure 1 shows the whole magnetization *curve* of the kagome antiferromagnet. Without ambiguity, one finds plateau structures at fractions, $M/M_{sat} = 0$, 1/9, 1/3, 5/9, and 7/9, as well as a jump from the 7/9-plateau to the saturation value at exactly $H_s/J=3.0$, where $M_{sat}$ is the full magnetization. To verify the accuracy of our grand canonical curve, we perform a set of conventional DMRG calculation both on a long cylinder with open ends (See supplementary Fig. S1), and on an open hexagonal cluster. By comparing these results, we confirm that the grand canonical analysis successfully gives the magnetization curve within the typical accuracy of $10^{-3}$ (see supplementary note 1).

**Singlet-triplet spin gap.**
The value of the singlet-triplet spin gap of the kagome antiferromagnet at zero field is still unsettled[8,9,22]. The evaluated spin gaps after the size scaling in the three latest DMRG studies are not fully consistent; as for the cylindrical-DMRG, Ref.9 has $\Delta=0.13(1)$ (the numerical data of Refs.8 and 9 at various fixed circumferences are basically consistent). There, the size scaling is basically given first along the long leg of a cylinder, and the extrapolation is given along the circumference of $L \leq 20$. As for Ref.22, they increased the size within $N=36-108$ by keeping the cluster to the square-like shape with periodic boundaries, and obtained $\Delta=0.055 \pm 0.005$. In fact, we checked these results carefully with the similar scaling, and found that the results depend much on the way the size scaling is performed[23] (see supplementary note1 & 2).

By contrast, in our grand canonical calculation, the size dependence becomes negligible (less than $10^{-3}$ in two dimension, see Methods) once we enter a cluster size of the proper system length. Therefore, one could evaluate the spin gap by the onset value of $H/J$ in the magnetization curve near zero field. In fig. 1 we find $\Delta = 0.05 \pm 0.02$ (see the red shaded region), which is obtained on a hexagonal cluster.

We briefly mention that our results are fully consistent with the data of the previous conventional DMRG studies: in our grand canonical DMRG on a cylinder with fixed small circumference (see supplementary Fig.S1), the spin gap gives more than twice as large values as the value mentioned above. This value should be compared with the data in Ref.9 on a long cylinder with the same circumference. For a proper extrapolation of the cylindrical results to a bulk two dimension, one needs to enlarge both the length and the circumferences simultaneously[23]. In fact, our grand canonical spin gap on a hexagonal cluster is very close to those of Ref.22 on a square cluster.

**Zero- and 1/9-plateaus.** The zero-plateau ranging at $0 \leq H/J \leq 0.05$ is the continuation of the zero-field ground state. Correspondingly, in our calculation, the spin structure in real space turned out to be completely structureless (See supplementary Fig. S2). One way to identify the nature of the spin liquid is to calculate the von Neumann entropy, $S = -\text{Tr}(\rho \ln \rho )$, defined on a subsystem of a long open cylinder by the conventional DMRG, where $\rho$ is the reduced density matrix of the subsystem. The value should follow, $S \sim \eta L_y - \gamma$, where $L_y$ is the circumference, $\eta$ is a constant, and $\gamma = \ln(D)$ is the topological entropy. In Refs.2 and 9, the topological dimension, $D$, of the ground state is given as $D \sim 2$, which supports the gapped $Z_2$ spin liquid.

In the 1/9-plateau state, the real space profile of the spin structure is rather intriguing; several geometries breaking the translational symmetry are quasi-degenerate (see supplementary Fig. S2 & note 3), and their stability is sensitive to the shape and size of the cluster. We consider this to be the good reason that the symmetry breaking long order is absent. Therefore, we perform the conventional DMRG and calculate the entanglement entropy of the 1/9-plateau state in the same manner as Refs.2 and 9, as shown in Fig.2; note that to have the 1/9-magnetization, we need to keep the system size the multiple of nine, and thus the choice of the clusters are limited compared to the calculation on the $M/N=0$-ground state. The topological dimension obtained in the $L_y=0$ limit seemingly gives the value $D=3$. Therefore, the spin gapped state at 1/9-magnetization is possibly *a $Z_3$ spin liquid, and is the first example of a spin liquid plateau induced by the magnetic field.* Even a $Z_3$ spin liquid itself has so far been observed only in a specified bosonic model[24], and the present model gives a more realistic setup. Further examination is required to identify the detailed nature of this phase.

**1/3-, 5/9- & 7/9-plateaus.** Contrasting to the first two plateaus, the rest of the plateaus have symmetry-breaking long range orders. Figures 3a-3c show the real space profiles of the magnetization density for 1/3, 5/9, and 7/9 plateaus. All of them are based on a same unit of a hexagram which holds nine lattice sites. This magnetic (extended) unit cell is three times as large as the original unit cell, namely, $Q_{mag}=3 \times 3=9$, with the spin density shown in Fig.3d. Such symmetry breaking requires strong interaction between boson, and the emergence of three such plateaus in a single system is already a quite unexpected matter to happen.

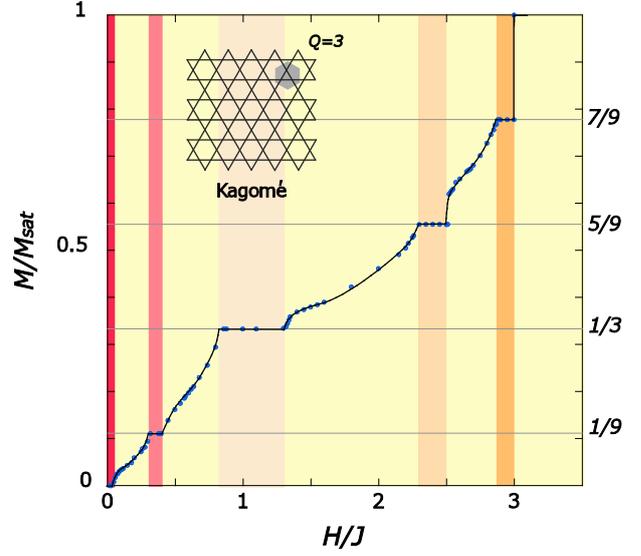

**Figure 1 | Magnetization curve of the spin-1/2 kagome Heisenberg antiferromagnet in a uniform magnetic field.** The saturation value of the magnetization density per site is $M_{sat}/N=1/2$. The inset shows the geometry of the kagome lattice. The shaded hexagon is the original lattice unit cell including three sites ($Q=3$). Data points are obtained by the Grand canonical analysis on a hexagonal cluster with $N=114$ and 132, which directly gives the curve of the thermodynamic limit without any size scaling. The range of each plateau is highlighted.

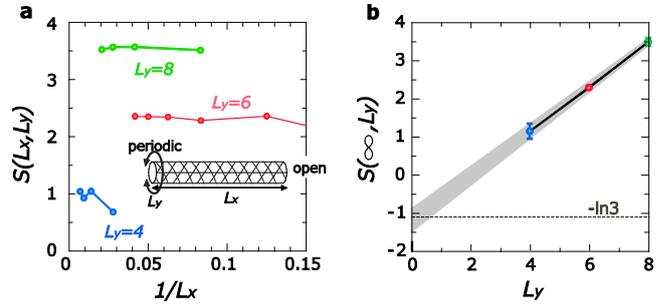

**Figure 2 | Entanglement entropy of a 1/9 magnetization plateau .** The results here are calculated on a long cylinder by the conventional DMRG. (a) $S(L_x, L_y)$ as a function of $1/L_x$ is given for $L_y$ =4,6,8, where $L_x$ and $L_y$ denote the number of sites along the leg and the circumference of the cylinder, respectively. (b) The value extrapolated for the infinite length, $L_x=\infty$, is given as a function of circumference $L_y$. The best fit to $S=\eta L_y - \gamma$ gives $\gamma=1.18 \pm 0.3$.

## Discussion

In spin-1/2 quantum magnets, a conventional (nontopological) nonmagnetic state basically comprises singlet, a unit of spin-0, often represented by the quantum fluctuation between two spins, $(|\uparrow\downarrow\rangle-|\downarrow\uparrow\rangle)/\sqrt{2}$. A breaking of singlet yields a bosonic elementary particle carrying spin-1, which is called magnon. The magnetic field controls the density of these bosons, serving as a chemical potential. As in the Mott insulator, there are particular values of the boson densities commensurate with the lattice periodicity[25], at which the gapped states are strongly pinned. At these fillings, a finite field range representing the spin gap is formed, which is the magnetization plateau.

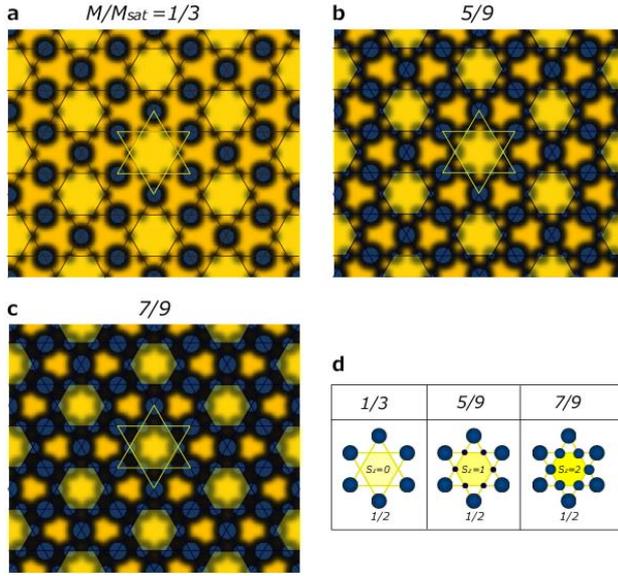

**Figure 3 | Density plot of the magnetization of the spin gapped states in 1/3, 5/9, and 7/9 plateaus (a)-(c).** The diameter of the blue circle on each lattice site scales the magnetization density. Hexagram consisting of nine sites indicate the magnetic unit cell, extended from the original one by three times ($Q_{mag}$=9). Hexagons in light-yellow are the guide to the eye. (d)Schematic alignment of spins on a hexagram, where the numbers indicate the magnetic density ($S_z$) on each vertices and on hexagons. Numerically, the exact magnetic density of the vertices of hexagram of (a) and (b) are shrunk due to quantum fluctuation by about 2-5% from the fractional values given in (d).

It is known that *the magnetization plateau emerges only when the quantity, $Q_{mag}S(1-M/M_{sat})$, is an integer* [26,27], where $Q_{mag}$ is the number of sites included in the unit period of the ground state, and $S$ is the spin quantum number which is 1/2 for the present case.

With this in mind, let us discuss the nature of our plateaus. We first examine the magnetic structure of $M/M_{sat}$ =0 and 1/9 plateaus, and find that they are possibly structureless in real space (see supplementary note3). Namely, the period of the ground state is the same as that of the lattice unit cell, $Q_{mag}=Q=3$, which gives, $Q_{mag}S(1-M/M_{sat})= 3/2$ and $4/3$, respectively. Since they are not integers but fractional numbers, the above conventional condition to have a plateau is not fulfilled. However, in two dimension, there is another way to form a spin gapped state (plateau) other than the above mentioned interplay with the lattice; it is to form *a structureless spin liquid*. The elementary excitation of such spin liquids by the magnetic field is no longer a magnon, but a deconfined spinon, carrying spin-1/2. While such exotic *spin liquid plateaus* could emerge *at a fractional value of $Q_{mag}S(1-M/M_{sat})$* as discussed in field theoretical study[28], it had been observed neither in theoretical models nor in materials. In fact, the calculations on the entanglement entropy indicate that zero-th and the 1/9-plateau form the spin liquid phases of topological dimension $D=2$ and $3$, respectively.

Contrastingly, in the latter three plateaus, we find $Q_{mag}S(1-M/M_{sat})$=3,2, and 1 (integers) for 1/3-, 5/9-, and 7/9-plateau, respectively, which all clearly fulfill the above conventional condition. Let us now discuss the origin of these solid plateaus.

In the 1/3-plateau, each triangular unit should hold net magnetization of 1/2, which consists of one up spin-1/2 and two spins forming a singlet (see Fig.4a). Similar to the zero-field Ising ground state, there are massive numbers of configuration of the 1/2-magnetized triangular unit[29], which is in fact a typical characteristic of the frustrated system. If these configurations are mixed up quantum mechanically, a liquid phase should emerge. To realize instead the solid state actually observed, one needs to select a particular configuration, and the problem reduces to how we pave this triangular unit on the kagome lattice to maximally gain energy.

In each configuration, one could draw a string along the singlet bonds of the triangular units as shown in the left panel of Fig.4a. Since every triangle shares its corners with the neighbouring triangles, the string never crosses with other strings, but continues until it meets itself again (otherwise it will extend toward infinity). In addition to the random configuration of strings, the representative two regular patterns are shown in Fig. 4a: a long string forming stripe and a shortest closed loop around the hexagon. One then needs to know which gains the energy, *the longer string or the shorter loop* due to the quantum mechanical resonance of spins along the string. The answer is the latter (see supplementary Fig.S3a)--- the kagome is fully tiled with hexagrams --- a symmetry breaking plaquette order is formed [30].

Once all the vertices of the hexagram (3 sites / 9 unit) are filled with a fully polarized up spin moment ($S_z$=1/2) at $M/M_{sat}$=1/3, a further simplified picture may work; focus on each hexagonal plaquette and isolate it by effectively neglecting the quantum fluctuation between the plaquette and the vertices of the Star, as shown in Fig.4b. This approximation is valid as far as the vertices of the hexagram are fully polarized. The interactions ($J S_z S_z$-term) between the plaquette and vertices work as an internal magnetic field, $H_{int}$=-$J$ per site on a plaquette. Figure 4c shows the magnetization process of the isolated plaquette in an effective field, $H+H_{int}$, i.e., the doping of magnons by the effective chemical potential. Each step of the big staircases corresponds to the increasing $S_z$ value or the number of magnons in the isolated plaquette. Now, notice that the point where the upshift of the staircases crosses the bulk magnetization curve coincides with the inflection point of the curve. This indicates the following scenario; If we condense the massive numbers of hexagrams, the quantum fluctuations between them become coherent throughout the system and works to destroy the staircases from the edge of toward the center of the step. The curve above/below the inflection point is the ruin of the edge of upper/lower staircase. This result, thus supports the picture that a hexagon works as a self-organized pseudo atomic orbital consisting of three discrete energy levels. Doping magnons to each level yields a series of plateaus starting from 1/3.

At present, the only other quantum magnet that possibly reveals comparably rich phase transitions is the $SrCu_2(BO_3)_2$[31,32]. However, the spin gapped phases of this material is based on a conventional singlet. In forming solids, they expand the unit cell to allocate the singlets in a regular period in a sea of doped magnons. Contrastingly, in our kagome, a non-trivial unit based on a hexagram is self-organized by the quantum many body effect. The doped magnons come into this cell in such a way that the electrons go into the quantum dots in an artificial semiconductor devise.

The above picture, then gives a strategy to design a system that could control the degree of frustration by the doping of particles; First, prepare an unfrustrated unit that could store several numbers of particles (in a kagome, this corresponding to a hexagon which could hold three magnons). Then connect them by the frustrated bonds. For example, this rule gives us a checkerboard lattice and its relevant pyroclore lattice. Their undoped ground states could be a spin liquid as far as the microscopic interactions are the finely balanced. We predict that dopings will also give rise to interesting states of matter as we find in kagome.

We further pay attention to the fact that the bose-Hubbard model at zero field[33] and the projected quantum dimer model[34], both on the same kagome lattice, are considered to have $Z_2$ liquids. Then, one may also expect the 5/9 and 7/9-plateaus to undergo a solid-to-spin-liquid transition[28] if the additional effect enhancing the quantum fluctuation is added to the Hamiltonian. The present findings will give some clue to find numbers of such exotic state of matter in kagome and in the above mentioned lattice models, many of which are still unexplored.

Regarding the laboratory systems of the kagome Heisenberg antiferromagnet, there are several candidates, e.g., $BaCu_3V_2O_8(OH)_2$[5], $ZnCu_3(OH)_6Cl_2$[6], $Cu_3V_2O_7(OH)_2 \cdot 2H_2O$[35], and $Rb_2Cu_3SnF_{12}$[36], while they are often distorted, anisotropic, or include complicated interactions, and the establishing of the ideal kagome Heisenberg material is long awaited. Our bulk magnetization curve is the only data numerically obtained in lattice models that could be compared directly and quantitatively with experimental measurements. Such comparison could be used to identify them as an ideal kagome. Also, if a 1/9-plateau were to appear in the experimental magnetization curve, one could examine, by neutron measurements or some other local magnetic probe like nuclear magnetic resonance, whether the state is structureless in reality. The magnetic specific heat and thermal conductivity measurements could further give us theorists a hint to know what kind of nonmagnetic as well as a magnetic elementary excitations are possible for this newly emergent phase. Thus our present findings would provide both theories and experiments a rich playground in search of open paradigms in quantum many-body physics.

each hexagonal plaquette unit. By putting in magnons from the empty level at 1/3-plateau, the 5/9, 7/9 and fully polarized states are formed. **(c)** Solid line is the magnetization step of the isolated hexagon in an effective field, $H+H_{int}$, where $H_{int}=-J$ is from the surrounding six vertex sites indicated in red circle. Broken line is the bulk magnetization curve of Fig.1. See Supplementary Fig.S3 & note 4 for more details.

## Methods

### Grand canonical analysis.
The technique is developed very recently by the authors[17,18], and gives the infinitesimally small response to the change in the external field. The physical quantities we get mimic their thermodynamic limit within the order of $10^{-4}$ for one dimension and $10^{-3}$ for two dimension, even in a relatively small cluster size of the order of $N \sim 10\text{-}100$. We briefly summarize the essential framework; the method smoothly divides the finite size cluster into the center part and the edges, and the main part reproduces the continuous bulk response by using the nearly zero energy edge state as a buffer. At a fixed system size and shape, we introduce the modulation of the energy scale by an externally given function, $f(r)$, at a location $r$, which smoothly deforms the Hamiltonian from the maximum at the center of the system ($r = 0$) to zero energy at the open cluster edges ($r=R$). After we obtain the proper eigen wave function of the deformed Hamiltonian, we evaluate the magnetization, $<S_z(r)>$. Now, remind that the total magnetization, $M_g = \Sigma\, S_z(r)$, of the deformed Hamiltonian is a conserved quantity given by hand. However, the expectation value of the local magnetic density $<S_z(r)>$ is no longer equal to $M_g/N$, but has a particular $r$-dependence: it takes nearly a uniform value at the center (though oscillating slightly, the mean values are uniform), then, often takes a peak or valley at the edges (see Suppplementary Fig. S4b & note 5). This is because the system optimizes the way wave function so as to realize the center value, $<S_z(r\sim 0)>$, to its thermodynamic limit, $m=MN$, at a given magnetic field. The excess/deficient magnetization, $M_g$-$M$, is absorbed/provided by the localized edge states of the cluster, which has the measure-zero energy due to $f(r=R) \sim 0$. These edges serve as a grand canonical bath.

The cluster and the function used in the present calculation are given in supplementary Fig. S4a. The numerical solver of the deformed Hamiltonian is not necessarily restricted to DMRG.

The reason why such mechanism works is discussed in detail in Ref.18 in the context of real space energy renormalization, together with applications to several two-dimensional quantum spin systems and to the electronic systems.

### Details of the DMRG calculation.
We have used the DMRG as a solver of the deformed Hamiltonian of the ground canonical analysis. The accuracy of the results is systematically controlled by the number of keeping basis state $m$ in the DMRG calculation and the typical error in the energy of the present calculations is around $10^{-4}$ for $m=4000$. As discussed in the Results, the aspect ratio closer to 1 gives the results closer to the bulk limit. Therefore, we adopt the hexagonal cluster of $N=114$ and $132$, instead of a long cylinder. More details including the numerical details of the results are available in the Supplementary Fig.S4 & note 5.

Regarding the calculation on the entanglement entropy in Fig.2, we used the conventional DMRG on a long cylinder with open edges and periodic circumferences. There, the number of states kept are $m=2000\text{-}10000$, and the data points displayed in Fig.2a are derived after the extrapolation to $1/m=0$.

### AcknowledgIments
We thank Keisuke Totsuka, Karlo Penc, Kenn Kubo, Hiroshi Kageyama, Yasu Takano, and Peter Fulde for comments and discussions. This work was supported by Grant-in-Aid for Scientific Research (No. 22014014, 20102008 & 25800204) from the Ministry of Education, Science, Sports and Culture of Japan.

### Author contributions
S.N. developed a code, carried out the DMRG simulations and analyzed the data with contributions from N.S. and C.H. C.H. brought up the project and wrote the paper discussing with N.S. and S.N. The project was

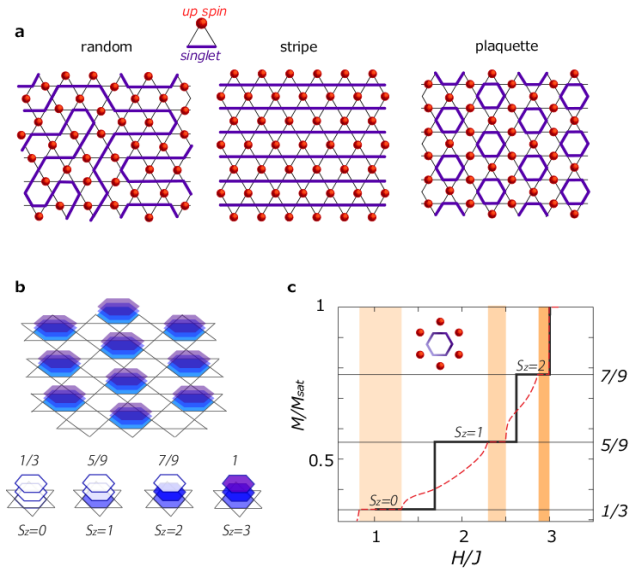

**Figure 4 | Details of the effective model describing the 1/3, 5/9, and 7/9 plateau states. (a)** A unit triangle at *1/3-plateau* includes one up spin (red circle) and two spins forming singlet bond (blue bond). The left panel gives one snapshot of random configuration of unit triangles, which includes winding singlet loops of several lengths. Other two panels are the regular configurations in which the singlet bonds form stripes and hexagonal plaquettes. The latter plaquette long range order is formed in the 1/3-plateau. **(b)** Schematic illustration of the three energy levels in

designed by all three authors and the three are equally responsible for the results.

Correspondence and requests to be addressed to C.H. (Email: chisa@cc.kyoto-su.ac.jp

**Competing finantial interests**
The authors declare no competing finantial interests.

# Controlling frustrated liquids and solids with an applied field in a kagome Heisenberg antiferromagnet

## Supplementary Information

### Satoshi Nishimoto[1], Naokazu Shibata[2] & Chisa Hotta[3]*

[1] Institute for Theoretical Solid State Physics, IFW Dresden, 01171 Dresden, Germany
[2] Department of Physics, Tohoku University, Sendai 980-8578, Japan
[3] Department of Physics, Faculty of Science, Kyoto Sangyo University, Kyoto 603-8555, Japan.

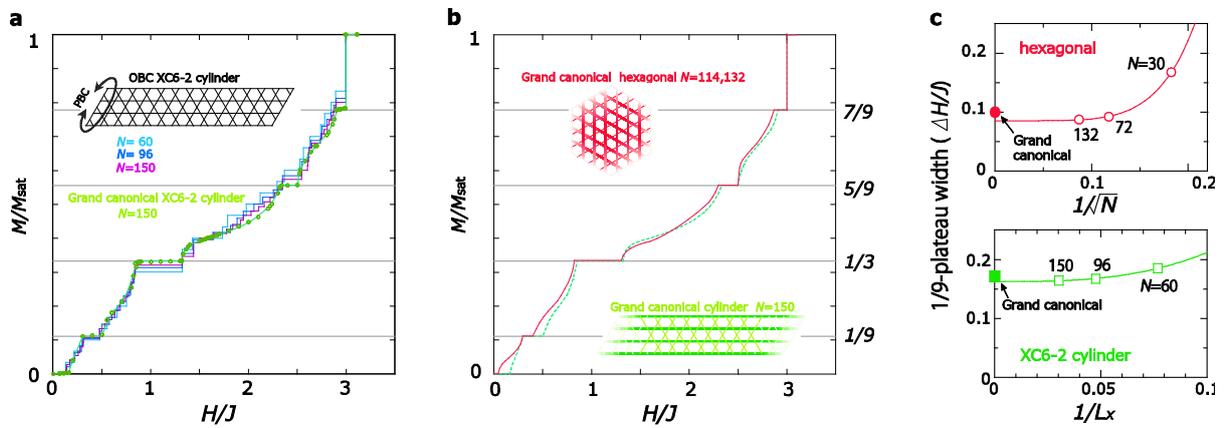

**Supplementary Figure S1 | Comparison of the magnetization curves.** (a) Comparison of the grand canonical curve (with data points) at $N=150$ and the staircases of the conventional open cylinder with $N=60$, 96, and 150, all on the XC6-2 cylinder. (b) Comparison of the two grand canonical curves using the hexagonal cluster with $N=114$ and 132 (red solid line, the same Fig. 1) and the XC6-2 cylinder(green broken line, the same as Fig.S1a), where the data points are abbreviated for simplicity. The name of the cylinder XC6-2 follows Ref.5. (c) System length dependence of the width of the 1/9-plateau of the conventional DMRG on the hexagonal (upper panel) and XC6-2 cylinder(lower panel), to be compared with the grand canonical ones plotted together on the same panels at $1/\sqrt{N}=0$ and $1/L_x=0$.

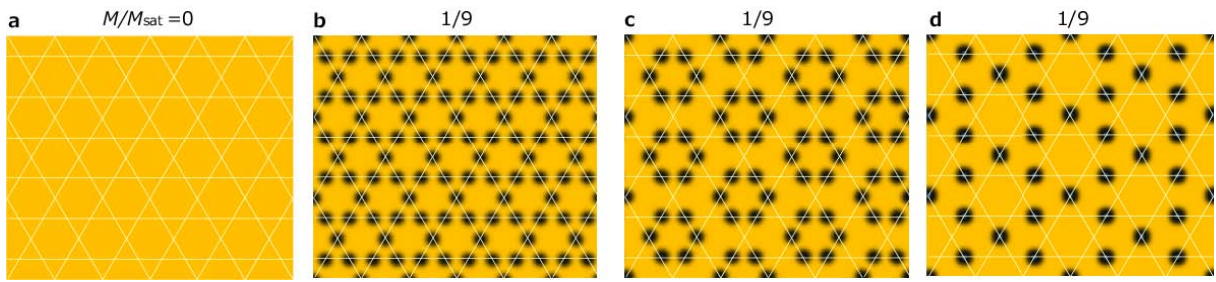

**Supplementary Figure S2 | Density plot of the magnetization of several spin gapped states of 0-th and 1/9-th plateaus.** (a) 0-th plateau is clearly structureless. In the 1/9-plateau, several regular configurations such as (b)-(d) are nearly degenerate. These 1/9-plateau structures obtained by the finite open clusters are numerically unstable, in contrast to the hexagram superstructures in the upper three plateaus (Fig.3, main text).

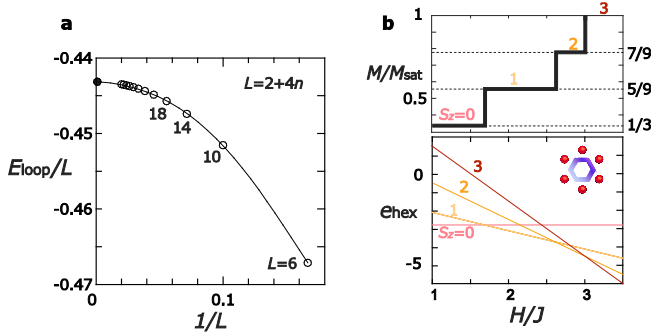

**Supplementary Figure S3 | Energy of the effective unit of solid plateaus.** (a) Energy per bond of the (isolated) spin-1/2 antiferromagnetic Heisenberg chain of length-$L$ with PBC. (b) Energy of a six-membered ring (hexagonal loop of spin-1/2 antiferromagnetic Heisenberg model), surrounded by the fully polarized six spins, $e_{hex} = \langle H_{hex}\rangle$ (see supplementary note 4). This unit gives the effective description of the hexagram in the plateau states, where total-$S_z$=0,1, and 2 of six spins correspond to the 1/3, 5/9, and 7/9 plateaus. A big magnetization staircase due to energy crossing, which is also given in Fig.4(main text), takes place at particular values of $H$, where the bulk magnetization curve takes an inflection point as in Fig.4.

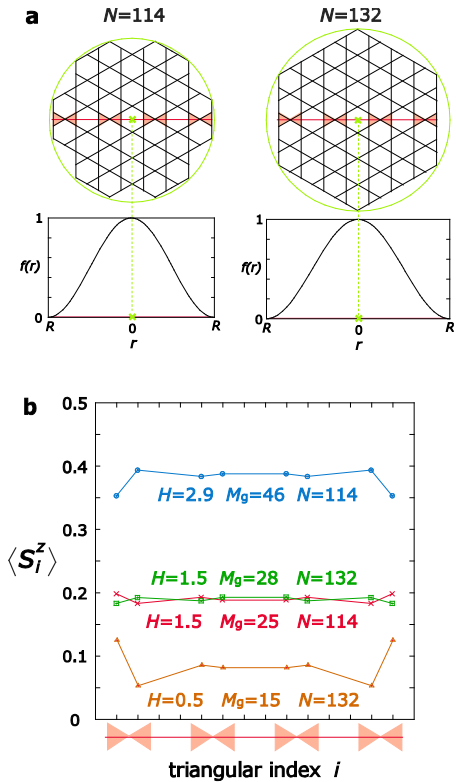

**Supplementary Figure S4 | Clusters used for the grand canonical analysis and the spatial profile of magnetization.** (a) Hexagonal clusters with $N$=114, 132 and the deformation function $f(r)=1-\cos(\pi r/R)$ on a polar coordinate with its origin at the center of the cluster. A large circle with radius $R$ represents the deformation range. (b) Expectation value of the local magnetic density $\langle S_z(r)\rangle$, averaged within each triangular unit shown in (a).

## Supplementary Note1: Comparison of numerical solutions with different conditions

As a reference to our main grand canonical DMRG solution on a hexagonal cluster in Fig.1, we prepared another series of solutions: a grand canonical DMRG solution on a long cylinder called XC6-2 (Refs.8 and 9) of $N$=150, and a set of conventional DMRG solution on the same XC6-2 cylinder of three different lengths, $N$=60, 96, and 150, with periodic boundary condition (PBC) along the circumference and the usual open boundary condition (OBC) at two open ends.

Since the latest DMRG calculations on the kagome antiferromagnet in Refs.8 and 9 were performed on several long cylinders with open edges, we tested our results first by comparing the ground state energy with the results in Ref.9; the ground state energy per site of the YC-10 cylinder [According to Ref.8, "YC9-2" denotes a cylinder (C) with some of the bonds oriented in the y direction (Y), with circumference of nine lattice spacings and a shift of two columns when connected periodically], after extrapolated to infinite length, is $e$= -0.4378(2) in Ref.9 and -0.437814 in our calculation, in almost perfect coincidence. The XC6-2 cylinder used in our calculation here gives the comparable value, $e$=-0.439205 in the infinite length limit. We adopted XC6-2 instead of YC-10 because its periodicity along the circumference is compatible with the magnetic structures of the plateaus we found. Otherwise, the results under small circumferences shall suffer a severe size effect.

Figure S1a shows the comparison of the grand canonical magnetization curve and conventional staircases on the XC6-2 cylinders. With increasing system length, the whole curve seems to approach the grand canonical one. This guarantees that our grand canonical DMRG actually gives the curve mimicking the ones in the thermodynamic limit. (See that the maximum-$N$ of the conventional DMRG reaches the same size as those of the grand canonical one, but still the staircase shows a finite size effect.). The magnetization values of 1/3, 5/9, 7/9-plateaus deviate from the exact rational values for finite $N$. This is because the cluster length and size is incompatible with the unit of the magnetic unit cell (hexagram) in these cylinders.

Next, in Fig. S1b, we compare the two grand canonical curves on a XC6-2 cylinder (the same as Fig. S1a) and on a hexagonal cluster (the same curve as Fig. 1 in the main text). Both curves clearly show five plateau structures we discussed in the main text. However, there exists a certain degree of deviation between the two curves at lower fields, $H/J$ <1. We adopt the one based on the hexagonal cluster as a true bulk curve. The reason is because the ones on a cylinder still needs to be extrapolated to infinite circumference, $L_y \to \infty$, in order to obtain the true thermodynamic limit in two dimension. Indeed, it was pointed out that taking a enough long circumference is often crucially important to obtain a result corresponding to the bulk limit[23]. However, the DMRG calculation on such large cluster is unavailable at present.

Contrastingly, if we take the isotropic hexagonal cluster and apply the grand canonical deformation equivalently to all the three directions (see below), the obtained results should directly be interpreted as those of the bulk limit even for a rather limited cluster size ($N$~100). There, the artifact of the OBC is suppressed due to the smooth radial edges. We already confirmed in Ref.18 that the grand canonical analysis on the isotropic cluster reproduces much better the one in the bulk limit than those on the cylinders in the case of a magnetization curve of a triangular lattice Heisenberg antiferromagnet.

## Supplementary Note2: Width of the plateaus.

In our grand canonical analysis, the magnetization value, $M/M_{sat}$, given in Fig. 1(main text) is basically almost independent of cluster size. The error against the thermodynamic limit is suppressed to less than the order of $10^{-3}$, once the side length of the cluster reaches the length, $L \sim 10\text{-}20$. Once we reach this accuracy, the size-dependence is negligible (see the last section of this Supplementary article).

In order to visualize this fact in the present model, we compare the width of the 1/9-plateaus of four data in Fig. S1a. As shown in the lower panel of Fig. S1c, the data of the conventional DMRG smoothly extrapolates to the value close to the grand canonical one at $L_x \to \infty$. In the upper panel, the similar behaviour is also seen for the hexagonal cluster with open edges. Thus, we can confirm that the grand canonical DMRG indeed gives the plateau width in the thermodynamic limit. Note that the extrapolated value for the XC6-2 cylinder is fairly lager than that for the hexagonal cluster because the size scaling in the $L_x$-direction is still missing for the cylinder.

## Supplementary Note3: Magnetic structures at $M_g/N=0$ and 1/9

We show in Fig. S2 the real space profile of the magnetization density of the low energy states at zero-th and 1/9-plateaus. These profiles are obtained by the calculation on finite size hexagonal clusters with open edges in order to generate, if any, a symmetry-breaking spatial structure. (The finite size cluster with periodic boundaries generally give only a uniform solution even when the superstructure is present in the bulk limit). We carefully compared the results (energies and profiles) with those of the other clusters as well.

In the zero-th plateau, all the sites remain $S_z=0$, with no structure observed, regardless of the shape and size of the cluster. This result is consistent with the scenario of a structureless spin liquid ground state reported so far by many groups.

As for the 1/9-plateau, there are quasi-degenerate manifolds of lowest energy states in a finite size cluster, and we show three of them with a regular spatial pattern in Figs. S2b-S2d: The first one is uniform, while the other two have the superstructure. All of them have very close energies. Further, these superstructures are very unstable against weak perturbations at the edges of the cluster. This may indicate that the magnetization density of the 1/9-plateau is also structureless in the thermodynamic limit. This is in sharp contrast to the case of the hexagram structure in the 1/3, 5/9, and 7/9-plateaus in Fig.3.

Therefore, it is likely that the 1/9-plateau also remains a liquid without any stable long range orders.

## Supplementary Note4: Effective model of 1/3, 5/9, and 7/9 plateaus.

In the Discussion(main text), we introduced an effective model to understand the characteristic magnetic structures in Fig.3 (main text) in unit of a hexagram. We show here a more detailed description to help the understanding.

In the 1/3-plateau state, the $N/3$-sites have fully polarized spins, $S^z_\mu = 1/2$, and the other $2N/3$-sites form closed loops, where each loop has total-$S^z = 0$. Since the interaction energy between the former and the latter do not depend on the configuration of the loops and spins, the energy difference among different configuration is determined solely by the summation of the energy of loops. A rough estimation of the energy of loops is given by regarding the loop of length-$L$ as an isolated antiferromagnetic Heisenberg chain of that length with PBC. Figure S3a shows the exact ground-state energy as a function of $1/L$, where $L=2+4n$ ($n \geq 1$: integer) is the length of the loop that may appear in the 1/3-plateau state of the kagome lattice. One finds that the energy gain per bond becomes smaller as the loop length increases. Thus, the lowest energy is obtained by keeping the loop to $L=6$ throughout the system, as shown in the right panel of Fig. 4 (main text).

In the following, we confine ourselves to the plaquette configuration in discussing the 1/3, 5/9, and 7/9-plateau states. Once the plaquette structure is formed, the energy is well-estimated in unit of an isolated hexagram, consisting of hexagon ($S_\nu$) and surrounding six vertex sites ($S_\mu$). Since the vertex spins are fully polarized ($S^z_\mu = 1/2$), the effective Hamiltonian of the hexagram is given by the hexagonal six sites as,

$$\mathcal{H}_{\text{hex}} = \sum_{\nu=1}^{6} S_\nu S_{\nu+1} - (H + H_{\text{int}}) \sum_{\nu=1}^{6} S_\nu$$

where $H_{\text{int}} = -J$ is the effective magnetic field due to $J S^z_\nu S^z_\mu$-term from up spins on the six vertex sites of the hexagram. Here, we neglect the $J(S^x_\nu S^x_\mu + S^y_\nu S^y_\mu)$-term because the vertex sites are assumed to be fully magnetized to $S^z_\mu = 1/2$. Actually, this quantum fluctuation slightly exists in the 1/3 and 5/9 plateaus on the kagome lattice, and $S^z_\mu$ on the vertex sites are reduced by a few percent.

By neglecting this fluctuation, the magnetization value allowed for this hexagon is total-$S^z = 0,1,2,3$. The exact energy of each the spin sector is given as a function of the magnetic field, $H$, in Fig. S3b. The transition between different spin sector takes place, and the large staircases of magnetization is formed (the upper panel of Fig.3b and Fig.4 in the main text).

## Supplementary Note5: Grand canonical Density Matrix Renormalization Group Method.

In the main text (Figure 1), we show the bulk magnetization curve of the kagome antiferromagnet obtained by the grand canonical analysis based on DMRG. This method allows for numerical determination of physical quantities mimicking their thermodynamic limit even in a relatively small cluster size of the order of $N \sim 10\text{-}100$. On a fixed cluster, we introduce a scaling function, $f(r)$, at a location $r$, which smoothly deforms the Hamiltonian from the maximum (~original Hamiltonian) at the center of the system to zero energy at the open cluster edges. We use the hexagonal cluster of $N=114$ and 132, shown in Fig. S4a, to obtain the magnetization curve in Fig. 1 (main text). The deformation function for these hexagonal cluster is $f(r)=(1+\cos(\pi r/R))/2$, where the constant $R$ is typically taken as the maximum distance between the center of the cluster and the edge sites. (For a grand canonical analysis on a long cylinder in Fig. S1, we interpret $r$ as a one dimensional coordinate along the leg of the cylinder, and take $f(r)$ uniform along the circumference direction.)

The eigen wave functions of that deformed Hamiltonian are obtained by the DMRG, and the expectation value of the local magnetization, $\langle S^z(r) \rangle$, is evaluated.

Some examples of the spatial dependence of $\langle S^z(r) \rangle$ is shown in Figure S4b as a longest cross section of the hexagonal cluster. The data in other directions not shown here are uniform, depending only on the distance, $r$, from the center.

The total magnetization, $M_g = \Sigma\, S_z(r)$, of the deformed Hamiltonian is a conserved quantity given by hand. However, the expectation value of the local magnetic density is no longer $M_g/N$, but shows particular special modulation: it takes nearly a uniform value at the center (though oscillating slightly, the mean values are uniform), then, often takes a peak or valley at the edges, depending on the choices of $M_g$. The center value, $\langle S^z(r \sim 0) \rangle$, gives the magnetization mimicking the one at the thermodynamic limit, $M/N$. If $M_g$ is larger/smaller than $M$, the excess/deficient magnetization, $M_g - M$, is absorbed/provided by the localized edge states of the cluster, which has the measure-zero energy due to $f(r=R) \sim 0$. These edges serve as a grand canonical bath.

In practice, the value of $M$ is evaluated as a mean value near the center of the cluster. As one finds from the comparison between the results with $N=114$ and $132$ for $H/J=1.5$ (they oscillate out of phase, with an almost identical oscillation center), the obtained $M=0.190$ is almost independent of the cluster size or given $M_g$. In the case of $H/J=2.9$, the value of $M_g =46$ is slightly smaller than the optimal one, thus, $\langle S^z(r) \rangle$ at $r \sim R$ is pulled out to compensate for that loss. Indeed, the typical accuracy of the grand canonical analysis falls within the order of $10^{-4}$ once the maximum system length reaches the order of $L \sim 30\text{-}50$[17,18]. In the present case with $L \sim 16$, the accuracy is $\sim 10^{-3}$.

In the present calculation, we kept $m \sim 4000$ density-matrix states in the renormalization procedure; even at the severest cases (at low magnetic fields), the error of $\langle S^z(r) \rangle$ around the center of system is less than $10^{-3}$, compared to the ones after the extrapolation, $1/m \to 0$. The typical error in the energy per site is about $10^{-4}$.

In this way, this method is useful to determine the magnetization curve of quantum 1D and 2D magnets, and a chemical potential curve of the electronic system as well. Particularly, in a frustrated system where the quantum Monte Carlo method is not applied, the method reaches numerically the bulk physical quantities in quantum many body system for the first time. In fact, the magnetization of another frustrated magnet, a $S=1/2$ triangular Heisenberg model with a 1/3-plateau, is determined for the first time as an unbiased solution[17,18].